\documentclass{article}

\usepackage{arxiv}

\usepackage{graphicx}
\usepackage{comment}

\usepackage{epstopdf}
\usepackage{amsmath}
\usepackage{natbib}
\usepackage{enumerate} 
\usepackage[normalem]{ulem}
\usepackage{soul}	
\soulregister\citep7
\soulregister\citet7

\usepackage{url}
\usepackage{xcolor} 
\usepackage{hyperref}
\usepackage{array,booktabs,arydshln}

\definecolor{DarkBlue2}{rgb}{0,0.08,0.50} 
\hypersetup{
    colorlinks = true, 
    linkcolor = DarkBlue2, 
    citecolor = DarkBlue2, 
    urlcolor = blue, 
    anchorcolor = red
}

\hypersetup{
pdftitle={Calibrating the Nelson-Siegel-Svensson Model by Genetic Algorithm},
pdfsubject={q-fin.CP},
pdfauthor={Asif Lakhany, Andrej Pintar, Amber Zhang},
pdfkeywords={Yield Curve Modelling, Nelson-Siegel Model, Svensson Model},
}

\newcommand\VRule[1][\arrayrulewidth]{\vrule width #1}
\title{Calibrating the Nelson-Siegel-Svensson Model by Genetic Algorithm}
\date{\today}

\author{
    \href{https://orcid.org/0000-0002-3509-230X}{\includegraphics[scale=0.06]{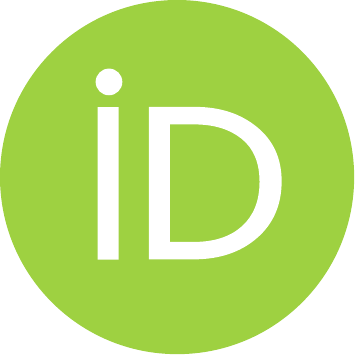}\hspace{1mm}Asif Lakhany} \\
    Markham, Ontario, Canada \\
    \texttt{Asif.Lakhany@gmail.com} \\
    \And
    \href{https://orcid.org/0000-0002-3384-8951}{\includegraphics[scale=0.06]{orcid.pdf}\hspace{1mm}Andrej Pintar} \\
    SS\&C Algorithmics \\
    Toronto, Canada \\
    \texttt{Andy.Pintar@sscinc.com}
    \And
    \href{https://orcid.org/0000-0003-0247-8107}{\includegraphics[scale=0.06]{orcid.pdf}\hspace{1mm}Amber Zhang} \\
    SS\&C Algorithmics \\
    Toronto, Canada \\
    \texttt{Amber.Zhang@sscinc.com} \\
}

\renewcommand{\headeright}{}
\renewcommand{\undertitle}{}

\begin{document}
\maketitle

\begin{abstract}
Accurately fitting the term structure of interest rates is critical to central banks and other
market participants. The Nelson-Siegel and Nelson-Siegel-Svensson models are probably the
best-known models for this purpose due to their intuitive appeal and simple representation.
However, this simplicity comes at a price. The difficulty in calibrating these models is twofold.
Firstly, the objective function being minimized during the calibration procedure is nonlinear
and has multiple local optima.
Secondly, there is strong co-dependence among the model parameters. As a result,
their estimated values behave erratically over time.
To avoid these problems, we apply a heuristic
optimization method, specifically the Genetic Algorithm approach,
and show that it is able to construct reliable interest
rate curves and stable model parameters over time, regardless of the shape of the curves.
\end{abstract}


\section{Introduction}
\label{sec:Introduction}

An accurate representation of the term structure of interest rates plays a key role
in the area of financial risk management and financial economics.
For example, investors often use the yield curve as a key representation
of future economic activity and inflation.
As pointed out in \citep{nielsen2017}, its slope indicates the short term interest rates
and is closely followed by investors. As such, estimation and representation of the
term structure of interest rates has been an active area of research for several years.

Smoothed bootstrap is one widely used method of approximating the term structure.
Practitioners use market data to bootstrap discrete spot rates and then use various spline
methods to fit a smooth and continuous curve from the set of discrete rates
\citep{mcculloch1971, mcculloch1975, vasicek1982, shea1984, steeley1991,
adams1994, fisher1994, waggoner1997}.
Because the smoothed bootstrap method has been criticized for having undesirable
economic properties or being a ``black box'' model \citep{seber2003},
\citet{nelson1987} proposed another parametric method in which the entire forward
curve is characterised by a single set of parameters. These parameters represent the long-run
level of interest rates, the slope of the curve, and humps in the curve. This
is a parsimonious model \citep{nelson1987}, and is consistent with the findings of
\citet{litterman1991} that most of the variation in returns on all fixed-income
securities represented by the yield curve can be explained in terms of three
attributes of the yield curve: level, steepness and curvature.
\citet{svensson1994, svensson1995} later extend \citet{nelson1987} by including
a second hump/trough factor to allow a more complex range of term structure shapes.

\citet{diebold2013} claim that the Nelson-Siegel (NS) and Nelson-Siegel-Svensson
(NSS) models are probably the best-known yield curve fitting tools among central
bankers, in industry, and in research. There are three main reasons:
\begin{enumerate}[i]
\item This model is able to construct yields for all maturities without using
interpolation,
\item Factors obtained from the NS model have meaningful economic interpretations,
\item Empirical test results show the calibrated yield curves fit the market
data reasonably well, especially in the long end of the term structure.
\end{enumerate}

First of all, NS and NSS models are extensively used by central banks and
monetary policy markers, due to their intuitive appeal and ease of implementation
\citep{boe2001, bis25, coroneo2011}.  Meanwhile, fixed-income portfolio
managers employ them to immunize their portfolios \citep{barrett1995}.  The NS and
NSS models also have recently experienced growing popularity in academic research.
\citet{boe2001, diebold2006, lorencic2016} use the NS model as the benchmark to judge
other models' performance in estimating and forecasting interest rate curves.
They find that the NS model appears to be much more accurate at long horizon
predictions than various standard benchmark forecasts. \citet{martellini2007}
employ the NS model to extract time-varying parameters used as proxies for
factors affecting the shape of the yield curve and observe that the NS model
achieves a high quality of fit with an average adjusted $R^2$ higher than 99\%,
and never below 99\%, in their estimation data. Furthermore, when these curve-affecting
factors are combined with a Student copula, \citet{martellini2007} are
able to generate Value-at-Risk (VaR) estimates that strongly dominate standard VaR
estimates in the out-of-sample tests. The NS model is also used to bootstrap the
riskless spot rate curves when calculating the credit risk spread
for the U.S. market \citep{xie2008}.  Finally, \citet{coroneo2011} conclude that the
NS yield curve model is compatible with the no-arbitrage constraints
on the U.S. market.

Although the NS and NSS models have many appealing features, the estimation of
model parameters is quite difficult. The estimates are usually obtained by
minimising the difference between the zero rates and the corresponding
observed market rates. This objective function is highly non-linear.
One common approach uses nonlinear regression techniques to obtain all
model parameters simultaneously. \citet{cairns2001} use the maximum likelihood
technique to fit the model and find that the time series of model parameter
estimates are very unstable. From one day to the next, the results can jump
from one location to another. They refer to this phenomenon as a catastrophic
jump: the size of the jump will typically be much larger than would be
consistent with the corresponding changes in price. As a result, the search
starting value is very important and the probability of getting local optima
for the next day is very high even if one uses today's global optimum.

Another approach used in practice is to linearize the nonlinear objective function
by fixing the shape parameter in the model that causes the nonlinearity, and
then to solve a much simpler problem. There are two common ways to conduct the
linearization. \citet{nelson1987} propose the ``grid search with OLS
regression'' method which is still one of the most popular market practices.
More specifically, they conduct the Ordinary Least Squares (OLS) regression to
estimate the remaining model parameters, conditional upon a grid of the fixed shape
parameter. The second branch of research focuses on identifying a good value
for the shape parameter and fixing it over the whole time series of term
structures.  For example, \citet{barrett1995, fabozzi2005} fix the shape
parameter at 3 for annualized returns.  \citet{diebold2006, coroneo2011} select
the fixed shape parameter where the maturity at which the maximized loading on
the medium-term is exactly 30 months to avoid hundreds of potentially changing
numerical optimizations under the grid search methodology.
In what follows, we will show that our methodology does not depend on any such ad-hoc procedures.

Although the linearized version of the model has a simpler objective function,
it is typically not convex and has many local optima. Therefore, classical techniques,
based on derivatives of the objective function, are not appropriate, and
can even generate negative long-term rates \citep{gilliGrosseSchumann2010, gilliSchumann2010}.
Regardless, researchers use such techniques.
\citet{boe2001, gurkaynak2007} and many more document that
the estimated parameters demonstrate instability, in that they sometimes jump
from one day to the next, often with little actual movement in
underlying bond prices. The instability of model parameter estimates also comes
from the selection of the fixed shape parameter.  \citet{dePooter2007} concludes
that the calibrated model parameters can take extreme values under different
fixed shape parameters. Therefore, it is critical to carefully determine the
shape parameter for each sample data and estimation window. Under the grid
search method, although no shape parameter is pre-defined, trying hundreds
or thousands of different starting points in the optimization is computationally
expensive.  However, the grid search method does not guarantee the
final solution is globally optimal. Additionally, the NS and NSS models are well
known for high multicollinearity between the slope and hump factor which can
inflate the variance of the estimators.  Consequently, they are more difficult to
calibrate \citep{cabrera2014, annaert2013}.

Because of these limitations in the traditional linear estimation of NS and NSS
model parameters, scholars are trying different approaches to deal with
the ill-conditioned optimization objective function.  \citet{annaert2013} propose
to switch to ridge regression whenever they detect high multicollinearity
under a normal grid search OLS regression procedure.  Although ridge
regression is able to improve the out-of-sample predictability at the short
and long ends of the term structure with lower mean absolute prediction errors and
to obtain a lower volatility for some NS model parameters than traditional grid
search, the shape parameter estimation still presents erratic time series
behaviour.  Meanwhile, the grid search method has to be conducted for every
observation date during the estimation period.  This is computationally expensive,
and the accuracy is limited by the size of the grid \citep{lakhany2000}.

In order to deal with the multiple local minima in the objective function, the
high-correlation problem among regressors, potential catastrophic
jump phenomena, and long computation time associated with the calibration of the NS
and NSS models, we apply a heuristic optimization method, or specifically a Genetic Algorithm
(GA) approach. We believe that GA can handle almost all of the above issues, due
to the following features \citep{goldberg1989}:

\begin{enumerate}[i]
\item Ability to handle non-differentiable, nonlinear and multimodal objective
functions.
\item Highly parallel implementation to cope with the computationally expensive
objective functions.
\item Ease of use; few control variables to steer the minimization. These
variables are robust and easy to choose.
\item Good convergence properties; ability to escape from local optima.
\item Good stability properties; ability to use today's solution as a good
starting value for the next observation day.
\end{enumerate}

By using the GA method, this paper finds that all model parameters estimated
on each observation date are consistent with their corresponding economic
meaning.

Our test results show that the NS yield curve has
difficulty fitting the entire term structure, especially the long-term
maturities of twenty years or more.  The reason for this is that the convexity of the yield curve
tends to pull down the yields on longer-term securities, giving the yield curve
a concave shape at longer maturities.  The NS specification, while
fitting shorter maturities quite well, tends to have the forward rates
asymptote too quickly to be able to capture the convexity effects at longer
maturities.  We find, however, that the NSS specification is
sufficiently rich to capture the double humps in the curve. On the other hand,
when the term structure of the calibrated yield curve shortens as the estimation
window moves forward, the NS model is preferable.

When calibrating NS and NSS model parameters for a group of yield
curve scenarios on a given date, we find that the two hump position parameters
($\lambda$,$\kappa$) are very similar across all scenarios. Therefore, we
only need to calibrate these two parameters once, and can re-use them to estimate
the remaining 3 or 4 factors for all remaining scenarios.  By representing the
scenarios in terms of the NSS parameters,
we are able to compress the risk factor space and represent the yield curve sensitivities
as a linear combination of the sensitivities on the NSS parameters.

The estimation of model parameters no longer shows erratic time series behaviours under the GA approach.
This demonstrates that the GA method makes the estimates more stable.
Furthermore, because today's solution is a good starting point for the next day's search, one can use
considerably fewer iterations to obtain suitable results.
This, combined with the fact that GA is well-suited to parallel computing,
makes it possible to obtain useful results in a
reasonable amount of time.  This is especially true for a long estimation window,
and when compared with the grid search method.

This paper makes several contributions to the interest rate calibration
literature. First of all, although the GA method has become more standard
in several fields of science, its application in estimation and modelling in
finance appears to still be limited. Our paper makes up for this gap.  Additionally,
we successfully solve the multiple-optima and multi-collinearity problem in the
calibration of the NS and NSS models and obtain all model parameters simultaneously.
Considering that the NS and NSS models are one of
the most popular interest rate models among market practitioners and
regulators, our paper paves the way for them to fully understand and predict
the interest rates and related derivatives.

The remainder of the paper proceeds as follows. Section \ref{sec:NSS} introduces the
NS and NSS models. Section \ref{sec:Calibrating} explains the GA technique used in this paper.
Section \ref{sec:DATA} presents the data and highlights relevant points.
Section \ref{sec:CALIBRES} illustrates the results and compares the model
results with the behaviour of market data.  It also examines through-time behaviour of the approach.
Section \ref{sec:Conclusions} concludes.

\section{Nelson-Siegel-Svensson Model}
\label{sec:NSS}
In this section we introduce the model. Our development closely follows \citet{annaert2013}.
In the Nelson-Siegel framework, the forward rate, $f(\tau)$ is modelled by the following expansion:
\begin{equation}
    f(\tau) = \beta_0 f_0(\tau) + \beta_1 f_1(\tau) + \beta_2 f_2(\tau)
\label{eq:ns}
\end{equation}
where $\lbrace \beta_i \rbrace {}_{i=1}^3$
are the coefficients of expansion and the basis functions $f_i(\tau)$
are given by:
\begin{equation}
\left.
\begin{aligned}
    f_0(\tau) &= 1 \\
    f_1(\tau) &= exp(-\tau/\lambda) \\
    f_2(\tau) &= (\tau/\lambda) exp(-\tau/\lambda)
\end{aligned}
\qquad\right\}
\label{eq:nsbasis}
\end{equation}
The basis functions are not arbitrary, but represent the basic components of an interest
rate term structure.
The first basis function represents the long-term level.
The second function represents an exponential decay and
allows the term structure to slope upwards (with $\beta_1 < 0$) or
downwards (with $\beta_1 > 0$). The third function produces a butterfly effect;
viz., $\beta_2 > 0$ will
produce a hump and $\beta_2 < 0$ will produce a trough. Finally, the parameter $\lambda$ determines
the location of the hump or the trough, as well as its steepness.

The interest rate term structure is obtained by averaging the forward rate curve
(\ref{eq:ns}) over different maturities. We will denote this curve by
$R(\tau) = (1/\tau) \int_0^\tau f(s) ds$. It is easy to check that
\begin{equation}
    R(\tau) = \beta_0 F_0(\tau) + \beta_1 F_1(\tau) + \beta_2 F_2(\tau)
\label{eq:nsi}
\end{equation}
where
\begin{equation}
\left.
\begin{aligned}
    F_0(\tau) &= 1 \\
    F_1(\tau) &= \frac{1 - exp(-\tau/\lambda) }{\tau/\lambda} \\
    F_2(\tau) &= \frac{1 - exp(-\tau/\lambda) }{\tau/\lambda} - exp(-\tau/\lambda).
\end{aligned}
\qquad\right\}
\label{eq:nsibasis}
\end{equation}
Equations (\ref{eq:nsi}) and (\ref{eq:nsibasis}) sufficiently describe term structures with a single
hump.  But in practical cases, market data may contain a second hump, generally at a longer
maturity, in form of a downward bend.  In order to model these cases, \citet{svensson1995} suggested using
a second hump/trough term. The Nelson-Siegel-Svensson model is described by:
\begin{equation}
    R(\tau) = \beta_0 F_0(\tau) + \beta_1 F_1(\tau) + \beta_2 F_2(\tau) + \beta_3 F_3(\tau)
\label{eq:nssi}
\end{equation}
where
\begin{equation}
\left.
\begin{aligned}
    F_0(\tau)   &  = 1 \\
    F_1(\tau)   &  = \frac{ 1 - exp(-\tau/\lambda) }{\tau/\lambda} \\
    F_2(\tau)   &  = \frac{1 - exp(-\tau/\lambda) }{\tau/\lambda} - exp(-\tau/\lambda) \\
    F_3(\tau)   &  = \frac{1 - exp(-\tau/\kappa) }{\tau/\kappa} - exp(-\tau/\kappa).
\end{aligned}
\qquad\right\}
\label{eq:nssibasis}
\end{equation}
Our main focus in this paper is to propose a robust and stable algorithm for calibrating the
coefficients
$\lbrace \beta_i \rbrace {}_{i=1}^3 \cup \lbrace \lambda \rbrace$
for the Nelson-Siegel model given by (\ref{eq:nsi}) and (\ref{eq:nsibasis}) or the set
$\lbrace \beta_i \rbrace {}_{i=1}^4 \cup \lbrace \lambda, \kappa \rbrace$
 for the Nelson-Siegel-Svensson model described by equations (\ref{eq:nssi}) and (\ref{eq:nssibasis}).
In what follows, we refer to this as the calibration problem.

\section{Calibrating the Nelson-Siegel(-Svensson) Model}
\label{sec:Calibrating}

In Section \ref{sec:NSS} we described the Nelson-Siegel and Nelson-Siegel-Svensson models.
We now propose methodologies for solving the calibration problem.
As mentioned before, each term in the NSS expansion represents a feature of the term structure, and,
as such, the parameters cannot take an arbitrary range. This means that fitting to the market
data is a bound constrained optimization problem. The method we propose due to its flexibility is
based on genetics and natural selection.  In this section we shall briefly touch upon the features
of the Genetic Algorithm (GA) approach.
Interested readers are referred to the comprehensive monograph by \citet{haupt2004}.
In short, GA comes with several features that are very suitable to the present application.
These include the ability to escape local minima, parallel
implementation, and the ability to
provide multiple solutions instead of a single local minimum.

We have implemented GA with the following features: we start with a randomly generated
population using Quasi Random sequence generator \citep{broda2016}.  In the case of NSS, this sequence is generated in
six dimensional space ($\lbrace \beta_i \rbrace {}_{i=1}^4, \lambda, \kappa$).
The first parameter, $N$, that we choose is the number of genes in the initial population.
We impose this number to be a multiple of 4, which facilitates partitioning the population
into winners and losers (half each), and parental gender (half again).
In case a repeated evaluation is desired under changing
market data we take $N=1,024$; for all other cases, $N=512$ should suffice.
We next define the fitness function which, simply put, is the negative
distance between the market data and the model values.  The algorithm goes through the fundamental steps of natural
selection, crossover and blending/mutation.  With this fitness function we are able to categorize the
population as elite, winners and losers.  The number of elite members
is taken to be three for these experiments.
The tournament to decide the mother and father of a newborn is also
based on three randomly selected winners of each gender.
Then we apply the crossover and blending techniques to obtain the newborn.
We use Eshelman-Shaffer Blending as described in \citet{haupt2004}.
In our implementation, the random weight used for blending
is generated so that the new point is never outside the bounds imposed by the problem. This
obviates the need for trial and error. Finally, we
apply mutations.  The mutation rate for
these experiments has been kept adaptive to the history of evolution. This idea is subject to criticism.
From a purist's approach, the need to adaptively modify the mutation rate may suggest that the GA is
not working. However, since our stopping criterion includes a limit on the number of generations,
we are unable to consistently obtain optimal results without incorporating this feature.
We claim that, with a sufficiently large number of generations, a constant mutation rate should produce
equivalent results.

From the practitioner's point of view, our GA implementation is equipped with the following three non-standard
features.  First, we improve the blending operation in such a manner that the resulting operation yields
parameter values within the original bounds constraints. For example, consider the following blending
of a parameter $p_i$ from the parents $p^{[MA]}_i$ and $p^{[PA]}_i$ (where the meaning of
\textit{PA}ter and \textit{MA}ter should be clear):
\begin{equation}
\left.
\begin{aligned}
    \hat{p}{}_i^1   &  = \mu p_i^{[MA]} + (1-\mu) p_i^{[PA]} \\
    \hat{p}{}_i^2   &  = \hat\mu p_i^{[MA]} + (1-\hat\mu) p_i^{[PA]}
\end{aligned}
\qquad\right\}
\label{eq:es}
\end{equation}
where $\hat{p}{}_i^j, j\in \{1,2\}$ is the new pair of blended values.
The numbers $\mu$ and $\hat\mu$ in formula (\ref{eq:es}) are generated from a uniform distribution $U(a,b)$.
In our implementation of this blending, we choose the parameters $a$ and $b$ in such a manner
that any number generated from this distribution will respect the bounds on the parameter $p_i$.
The next feature is the adaptive mutation feature discussed above.
Another feature, that we empirically justify, is skewing the initial population with a selected group
of genes.  We use this feature to address stability of the NS(S) parameters.
The keen observer will note that the hypothesis of the convergence
proofs that rely on uniform distribution of the starting population is violated \citep{haupt2004}.
Practically, the induced set of genes has cardinality much smaller (about 2\%-5\%)
than the population size. For example,
in our tests we use 16 returning winners out of a population size of 1,024 or 512.

\section{Data for testing the calibration fit}
\label{sec:DATA}
The set of data that we have used to test our proposed algorithm are OIS rates for the EUR currency taken
from two different periods.  By OIS rates we mean rates that were bootstrapped from
Overnight Index Swap (OIS) pricing data.
An OIS is an interest rate swap that exchanges a fixed rate payment with a floating rate payment on a fixed
notional amount for the life of the swap. In this case, the floating rate is based on a specific published
index of daily overnight rates in the currency of the swap. The term for the OIS typically ranges from 1 week to 2 years,
however longer-termed maturities on the order of 10 years are becoming more common \citep{hull2012}.
At maturity, the parties agree to exchange an amount, based on the agreed notional, as a difference of accruals from the
fixed rate and the geometric average of the floating index rate. Since there is no exchange of notional,
these swaps are virtually devoid of credit risk. The index rates used are computed differently for different
markets. For the U.S. market it is the daily effective federal funds rate.

In Table (\ref{table:oisRates})\footnote{
This data is sourced from Reuters, and is bootstrapped in conjunction with cubic spline
interpolation of the continuously compounded rate.}
we provide data for the dates September \{22, 23, 27, 28, 29, 30\}, 2011
(note the weekend is excluded), for the EUR OIS curve.
The first analysis is done on the data corresponding to September 22.
We selected this curve due to the existence of two humps.
It will allow us to demonstrate the applicability of the NSS model.
It was also chosen because of its complicated shape. In fact, careful observation suggests the presence of a
third hump at the end of the term structure.  In the settings of the algorithm we run, we shall use bounds that will avoid
capturing this third hump. Strictly
speaking, the NSS model itself is only meant to capture two such features.

Simply producing a good fit to complicated data is not the only objective of this paper.
As mentioned in the introduction, one of the biggest challenges associated with the NSS model is the
stability of the parameters under daily changing market conditions. To test how well
our algorithm handles the stability issue, we iteratively fit to a full week of data, provided in
Table (\ref{table:oisRates}).

\begin{table}[tbp]
\caption{EUR OIS Curve : September 22-30, 2011}
\centering
\begin{tabular}{!{\VRule} c !{\VRule} c !{\VRule} c !{\VRule} c !{\VRule} c !{\VRule} c !{\VRule} c !{\VRule} c !{\VRule} }
\hline
{}	&	2011-09-22	&	2011-09-23	&	2011-09-26	&	2011-09-27	&	2011-09-28	&	2011-09-29	&	2011-09-30	\\
\hline
Term	&	Rate \%	&	Rate \%	&	Rate \%	&	Rate \%	&	Rate \%	&	Rate \%	&	Rate \%	\\
\hline
30	&	0.9231	&	0.8910	&	0.8928	&	0.9187	&	0.9470	&	0.8896	&	0.8909	\\
60	&	0.7720	&	0.7328	&	0.7461	&	0.8125	&	0.8405	&	0.7959	&	0.8104	\\
90	&	0.7105	&	0.6726	&	0.6842	&	0.7552	&	0.7843	&	0.7624	&	0.7655	\\
180	&	0.5995	&	0.5793	&	0.6073	&	0.7012	&	0.7265	&	0.7005	&	0.6656	\\
270	&	0.5710	&	0.5481	&	0.5807	&	0.6859	&	0.7193	&	0.7033	&	0.6652	\\
365	&	0.5685	&	0.5417	&	0.5783	&	0.6815	&	0.7219	&	0.7135	&	0.6662	\\
455	&	0.5743	&	0.5465	&	0.5864	&	0.6852	&	0.7338	&	0.7288	&	0.6784	\\
545	&	0.5879	&	0.5612	&	0.6049	&	0.6969	&	0.7481	&	0.7477	&	0.6953	\\
635	&	0.6127	&	0.5860	&	0.6342	&	0.7209	&	0.7756	&	0.7768	&	0.7192	\\
730	&	0.6446	&	0.6211	&	0.6734	&	0.7550	&	0.8114	&	0.8110	&	0.7513	\\
820	&	0.6840	&	0.6665	&	0.7149	&	0.7945	&	0.8538	&	0.8501	&	0.7883	\\
910	&	0.7277	&	0.7158	&	0.7622	&	0.8412	&	0.8985	&	0.8953	&	0.8296	\\
1,000	&	0.7786	&	0.7710	&	0.8187	&	0.8941	&	0.9492	&	0.9478	&	0.8790	\\
1,095	&	0.8341	&	0.8313	&	0.8769	&	0.9513	&	1.0061	&	1.0062	&	0.9353	\\
1,185	&	0.8944	&	0.8927	&	0.9371	&	1.0114	&	1.0686	&	1.0703	&	0.9940	\\
1,275	&	0.9519	&	0.9534	&	0.9983	&	1.0732	&	1.1326	&	1.1352	&	1.0557	\\
1,365	&	1.0138	&	1.0195	&	1.0606	&	1.1364	&	1.1995	&	1.1982	&	1.1181	\\
1,460	&	1.0765	&	1.0832	&	1.1232	&	1.2005	&	1.2670	&	1.2660	&	1.1815	\\
1,550	&	1.1358	&	1.1458	&	1.1852	&	1.2623	&	1.3330	&	1.3304	&	1.2443	\\
1,640	&	1.1947	&	1.2080	&	1.2473	&	1.3251	&	1.3969	&	1.3936	&	1.3064	\\
1,730	&	1.2505	&	1.2669	&	1.3070	&	1.3857	&	1.4585	&	1.4547	&	1.3665	\\
1,825	&	1.3052	&	1.3252	&	1.3652	&	1.4440	&	1.5182	&	1.5134	&	1.4250	\\
1,915	&	1.3586	&	1.3804	&	1.4202	&	1.4989	&	1.5743	&	1.5701	&	1.4806	\\
2,005	&	1.4085	&	1.4325	&	1.4718	&	1.5505	&	1.6274	&	1.6228	&	1.5322	\\
2,095	&	1.4574	&	1.4839	&	1.5228	&	1.6015	&	1.6787	&	1.6743	&	1.5826	\\
2,190	&	1.5045	&	1.5330	&	1.5712	&	1.6503	&	1.7274	&	1.7242	&	1.6312	\\
2,280	&	1.5489	&	1.5791	&	1.6167	&	1.6964	&	1.7737	&	1.7716	&	1.6774	\\
2,370	&	1.5892	&	1.6220	&	1.6591	&	1.7396	&	1.8168	&	1.8155	&	1.7202	\\
2,460	&	1.6294	&	1.6640	&	1.7011	&	1.7822	&	1.8594	&	1.8585	&	1.7620	\\
2,555	&	1.6676	&	1.7038	&	1.7411	&	1.8222	&	1.8997	&	1.8999	&	1.8023	\\
2,645	&	1.7037	&	1.7407	&	1.7801	&	1.8612	&	1.9391	&	1.9393	&	1.8404	\\
2,735	&	1.7367	&	1.7749	&	1.8145	&	1.8955	&	1.9740	&	1.9756	&	1.8756	\\
2,825	&	1.7697	&	1.8086	&	1.8488	&	1.9298	&	2.0090	&	2.0111	&	1.9099	\\
2,920	&	1.8011	&	1.8407	&	1.8822	&	1.9636	&	2.0436	&	2.0452	&	1.9428	\\
3,100	&	1.8591	&	1.9002	&	1.9415	&	2.0251	&	2.1065	&	2.1082	&	2.0045	\\
3,285	&	1.9143	&	1.9559	&	1.9966	&	2.0831	&	2.1647	&	2.1670	&	2.0606	\\
3,465	&	1.9646	&	2.0066	&	2.0493	&	2.1371	&	2.2182	&	2.2213	&	2.1135	\\
3,650	&	2.0153	&	2.0558	&	2.1003	&	2.1893	&	2.2700	&	2.2731	&	2.1642	\\
4,380	&	2.1992	&	2.2377	&	2.2837	&	2.3744	&	2.4636	&	2.4678	&	2.3539	\\
5,475	&	2.3927	&	2.4252	&	2.4626	&	2.5574	&	2.6534	&	2.6602	&	2.5420	\\
7,300	&	2.4628	&	2.4788	&	2.5164	&	2.6236	&	2.7240	&	2.7330	&	2.6156	\\
9,125	&	2.4241	&	2.4419	&	2.4725	&	2.5820	&	2.6753	&	2.6799	&	2.5558	\\
10,950	&	2.3637	&	2.3873	&	2.4207	&	2.5314	&	2.6193	&	2.6182	&	2.4905	\\
14,600	&	2.3514	&	2.3672	&	2.4080	&	2.5220	&	2.6163	&	2.6110	&	2.4864	\\
18,250	&	2.4092	&	2.4278	&	2.4633	&	2.5743	&	2.6732	&	2.6646	&	2.5443	\\
\hline
\end{tabular}
\label{table:oisRates}
\end{table}

The OIS data described above has clearly been processed.
In order to show that our methodology works equally well for raw data,
we will apply our proposed algorithm to the data used to build USD yield
curves.\footnote{This data is sourced from Bloomberg}
The following groups of instruments are chosen:

\begin{enumerate}[i]
\item The first group of instruments is used to build the short end of the curve.
    These are typically the cash/deposit securities available for up to one year tenor.
    However, they are commonly used for curve construction for up to three or six months.
    We recommend including only a single cash instrument with tenor equal to that of the curve being built.
\item The second group of instruments is used to build the curve beyond the term points
    covered by the first group of instruments. This group contains USD coupon bonds.
\end{enumerate}

Our data is based on all
available USD (zero-)coupon bond yields trading on July 28, 2020 -- a total of 357
instruments. Based on a typical use case, we selected 1, 2, 3, 6, 9-month, and 1, 1.5, 2, 2.5, 3, 3.5, 4, 4.5, 5, 5.5, 6, 6.5, 7, 7.5,
8, 8.5, 9, 9.5, 10, 12, 15, 20, 25, 30-year term nodes. From these 357 bonds, we finally choose 29 instruments whose
maturities are closest to our selected term nodes. The list of bonds appear in Table (\ref{table:usdRates}).

\begin{table}[tbp]
\caption{USD Bonds}
\centering
\begin{tabular}{!{\VRule} c !{\VRule} c !{\VRule} c !{\VRule} c !{\VRule} c !{\VRule} c !{\VRule}}
\hline
Cusip	&	Coupon	&	Maturity Date	&	BidYield	&	MidYield	&	Issue Date	\\
\hline
912796XG9	&	0	&	2020-08-27	&	0.099132786	&	0.090235681	&	2020-02-27	\\
9127962H1	&	0	&	2020-09-24	&	0.106476035	&	0.100137247	&	2020-03-26	\\
9127962T5	&	0	&	2020-10-29	&	0.111559138	&	0.106486972	&	2020-04-30	\\
912796UC1	&	0	&	2021-01-28	&	0.124619268	&	0.119529857	&	2020-01-30	\\
9127962Q1	&	0	&	2021-04-22	&	0.131905229	&	0.130635968	&	2020-04-23	\\
9127963S6	&	0	&	2021-07-15	&	0.13447065	&	0.129391733	&	2020-07-16	\\
912828Z60	&	1.375	&	2022-01-31	&	0.1561949	&	0.1459276	&	2020-01-31	\\
9128282P4	&	1.875	&	2022-07-31	&	0.1594827	&	0.14802215	&	2017-07-31	\\
9128283U2	&	2.375	&	2023-01-31	&	0.1649988	&	0.15592475	&	2018-01-31	\\
912828S92	&	1.25	&	2023-07-31	&	0.1706244	&	0.1603927	&	2016-08-01	\\
9128285Z9	&	2.5	&	2024-01-31	&	0.1954981	&	0.18908325	&	2019-01-31	\\
912828Y87	&	1.75	&	2024-07-31	&	0.2134859	&	0.20969775	&	2019-07-31	\\
912828Z52	&	1.375	&	2025-01-31	&	0.2410912	&	0.23515535	&	2020-01-31	\\
912828Y79	&	2.875	&	2025-07-31	&	0.272484	&	0.2680819	&	2018-07-31	\\
9128286A3	&	2.625	&	2026-01-31	&	0.3144243	&	0.309736	&	2019-01-31	\\
912828Y95	&	1.875	&	2026-07-31	&	0.358822	&	0.3550592	&	2019-07-31	\\
912828Z78	&	1.5	&	2027-01-31	&	0.3991395	&	0.3956219	&	2020-01-31	\\
91282CAD3	&	0.375	&	2027-07-31	&	0.4339846	&	0.43171135	&	2020-07-31	\\
9128283W8	&	2.75	&	2028-02-15	&	0.4596749	&	0.45677725	&	2018-02-15	\\
9128284N7	&	2.875	&	2028-05-15	&	0.4735112	&	0.4697927	&	2018-05-15	\\
9128284V9	&	2.875	&	2028-08-15	&	0.4898128	&	0.4862126	&	2018-08-15	\\
9128286B1	&	2.625	&	2029-02-15	&	0.5144641	&	0.51104645	&	2019-02-15	\\
912828YB0	&	1.625	&	2029-08-15	&	0.5303851	&	0.52870125	&	2019-08-15	\\
912828Z94	&	1.5	&	2030-02-15	&	0.5543428	&	0.5527359	&	2020-02-18	\\
912828ZQ6	&	0.625	&	2030-05-15	&	0.5773463	&	0.57652695	&	2020-05-15	\\
912810FM5	&	6.25	&	2030-05-15	&	0.5428465	&	0.52192415	&	2000-02-15	\\
912810FP8	&	5.375	&	2031-02-15	&	0.5607927	&	0.5487083	&	2001-02-15	\\
912810FT0	&	4.5	&	2036-02-15	&	0.7299671	&	0.7275026	&	2006-02-15	\\
912810QK7	&	3.875	&	2040-08-15	&	0.9737774	&	0.9697724	&	2010-08-16	\\
912810RN0	&	2.875	&	2045-08-15	&	1.1736387	&	1.1715444	&	2015-08-17	\\
912810SN9	&	1.25	&	2050-08-15	&	1.2167632	&	1.2161395	&	2020-05-15	\\
\hline
\end{tabular}
\label{table:usdRates}
\end{table}

\section{Calibration Results}
\label{sec:CALIBRES}
A fundamental reason for choosing GA is due to the ease of doing bound-constrained optimization.
As mentioned in Section (\ref{sec:NSS}), the parameters of the NS(S) model are not arbitrary. They represent different
features of the yield curve that are observed in the market.  We can leverage this fact to provide
consistent bounds to the parameters of the NS(S) model during a GA run.
For example, based on economic intuition,
$\beta_0$ is close to the empirical long-term spot rate and, as such,
cannot be negative for the data we are analyzing.
We also leverage the fact that $\lambda, \kappa$ are the range parameters.
Based on this, we select the bounds that appear in Table (\ref{table:oisBounds}).

\begin{table}[tbp]
\caption{Bounds Constraints for the NSS Model - OIS Curve}
\centering
\begin{tabular}{!{\VRule} c !{\VRule} c !{\VRule} c !{\VRule} }
\hline
Parameter & Lower Bound & Upper Bound \\
\hline
$\beta_0$ & 0.0 & 0.10 \\
$\beta_1$ & -0.10 & 1.0 \\
$\beta_2$ & -2.0 & 2.0 \\
$\beta_3$ & 0.0 & 2.0 \\
$\lambda$ & 0.0 & 4.0 \\
$\kappa$ & 4.0 & 30.0 \\
\hline
\end{tabular}
\label{table:oisBounds}
\end{table}

Next we discuss the parameters of the algorithm. Any algorithm that is affected by a large number of parameters
is often criticized as being difficult to use. Practitioners may be indifferent to the selection of parameters: how they
affect the solution and what the optimal values are.  GA is no different in this regard.
As such, our policy was to fix the parameters based on the recommendations made in \citet{haupt2004}.
In Table (\ref{table:parameters}) we present those parameters.

\begin{table}[tbp]
\caption{Parameters for GA Experiments}
\centering
\begin{tabular}{!{\VRule} l !{\VRule} c !{\VRule} }
\hline
Parameter & Value \\
\hline
Random Number Generator & Sobol Quasi Random Generator \\
Minumum Population Size & 512 \\
Maximum Population Size & 1,024 \\
Number of Elite Genes & 3 \\
Tournament Size & 3 \\
Minimum Mutation Rate & 0.2 \\
Maximum Mutation Rate & 0.5 \\
Returning Genes & 64 \\
Stopping Criterion & Maximum Iterations \\
\hline
\end{tabular}
\label{table:parameters}
\end{table}

It should be noted that even though our implementation of the algorithm allows for more complex stopping criteria than
shown in Table (\ref{table:parameters}), for the purpose of this study we imposed a fixed number of generations.
A fixed number of generations makes it convenient to compare different runs of the algorithm.
In practice, however, one would choose a more suitable stopping criteria such as those discussed in \citet{haupt2004}.
Table (\ref{table:parameters}) indicates that the number of returning genes is set to 64. A smaller value could also suffice;
these are the genes that move from one time step to another.

\begin{figure}[tbp]
    \centering
    \caption{Fitting OIS Curve using Nelson Siegel Model}
    \includegraphics[width=0.9\textwidth]{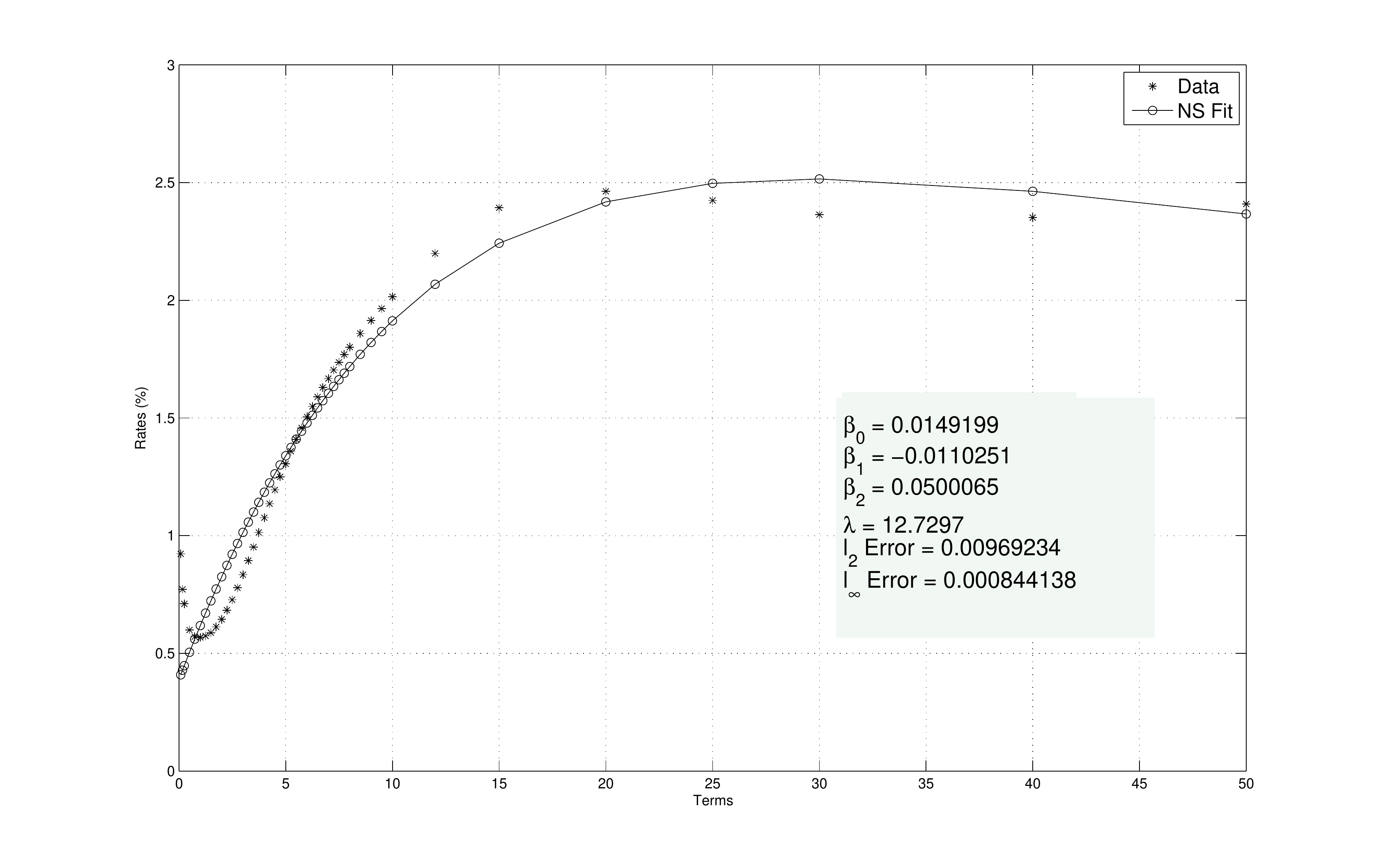}
    \label{fig:ns}
\end{figure}

Before we discuss the experiments to fit the OIS Curve presented in Table (\ref{table:oisRates}), we
need to demonstrate and empirically prove an earlier claim.
Our choice of data was driven by the fact that it exhibits three turning points - one of which at
the rear end will be ignored. Due to this, the NS model will be limited in its ability to fit the data.
Our first result in Figure (\ref{fig:ns}) shows exactly this. Note that the bound for $\lambda$ was taken
to be the union of the bounds for $\lambda$ and $\kappa$ in Table (\ref{table:oisBounds}).

\begin{figure}[tbp]
    \centering
    \caption{Fitting OIS Curve using Nelson Siegel Svensson Model, Population Size = 512}
    \includegraphics[width=0.9\textwidth]{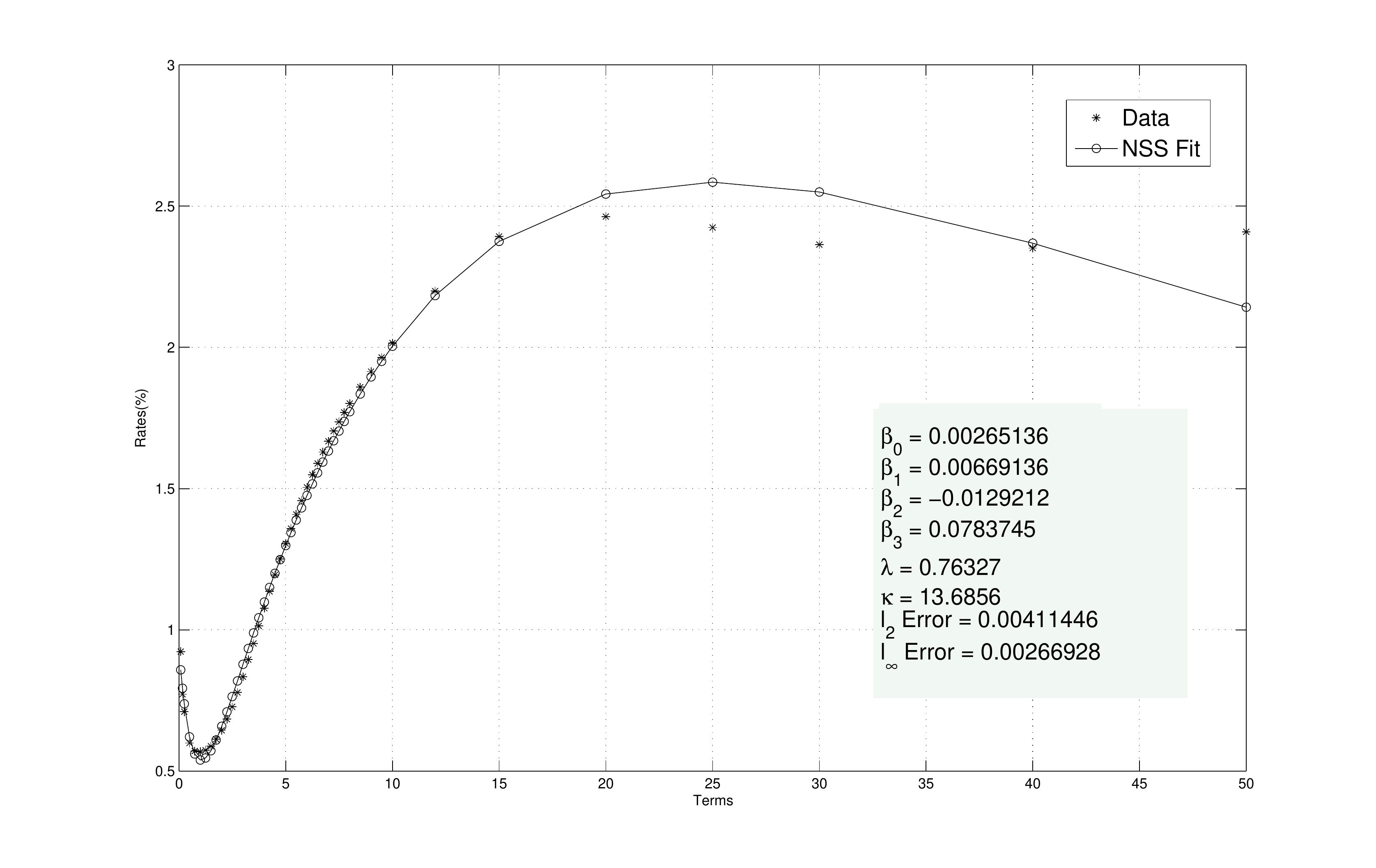}
    \label{fig:nssA}
\end{figure}

Now that we have confirmed the inadequacy of the NS model to handle the
September 22, 2011 data, we examine how well the NSS model handles it.
In Figure (\ref{fig:nssA}) we show the fit using the initial population
of 512 genes. We let the algorithm run for a total of 5,000 generations.

\begin{figure}[tbp]
    \centering
    \caption{Fitting OIS Curve using Nelson Siegel Svensson Model, Population Size = 1,024}
    \includegraphics[width=0.9\textwidth]{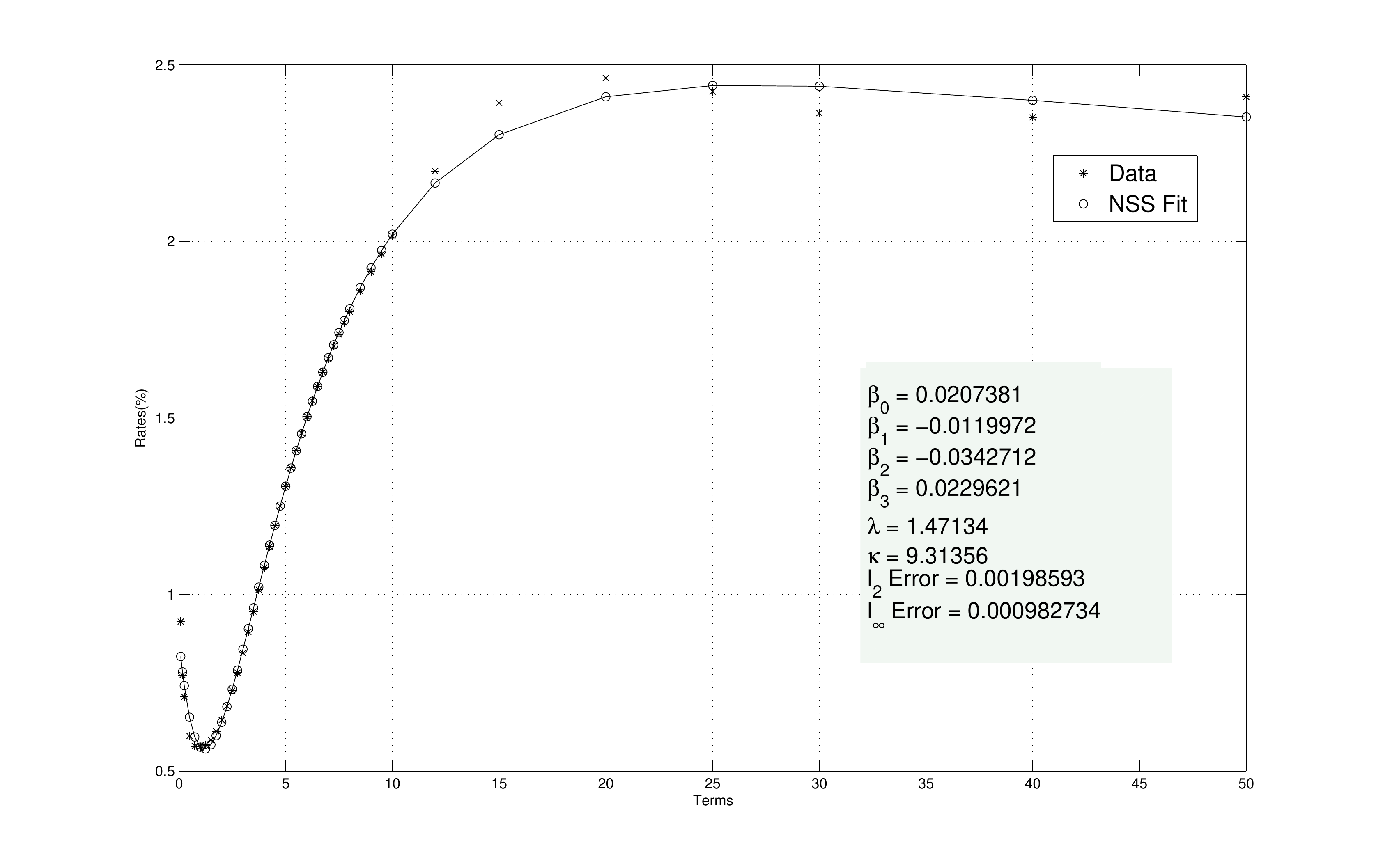}
    \label{fig:nssB}
\end{figure}

We then repeat the experiment with a population size of 1,024. We immediately notice a significant
improvement in the results for the same number of fixed iterations. However, we also see evidence
of the difficulty with the NSS model mentioned in Section (\ref{sec:Introduction}) and reported by
several researchers and practitioners. The change in the error is significant but not drastic; however, the
difference in the set of parameters produced by these two runs is substantial. This tells
us that the objective function admits several local minima.
Furthermore, it also shows why practitioners have had difficulty in obtaining a stable set of
parameters over time. The key requirement here is that, as data changes from day to day in a smooth manner,
the parameters of the NSS model should change proportionately.
This is the basic requirement of a stable algorithm.
Drastic changes in the calibrated parameter set
with only a small changes in the data make the fitting technique unreliable.

\begin{figure}[tbp]
    \centering
    \caption{Daily Movements in OIS Curve During the Period September 22-30, 2011}
    \includegraphics[width=0.9\textwidth]{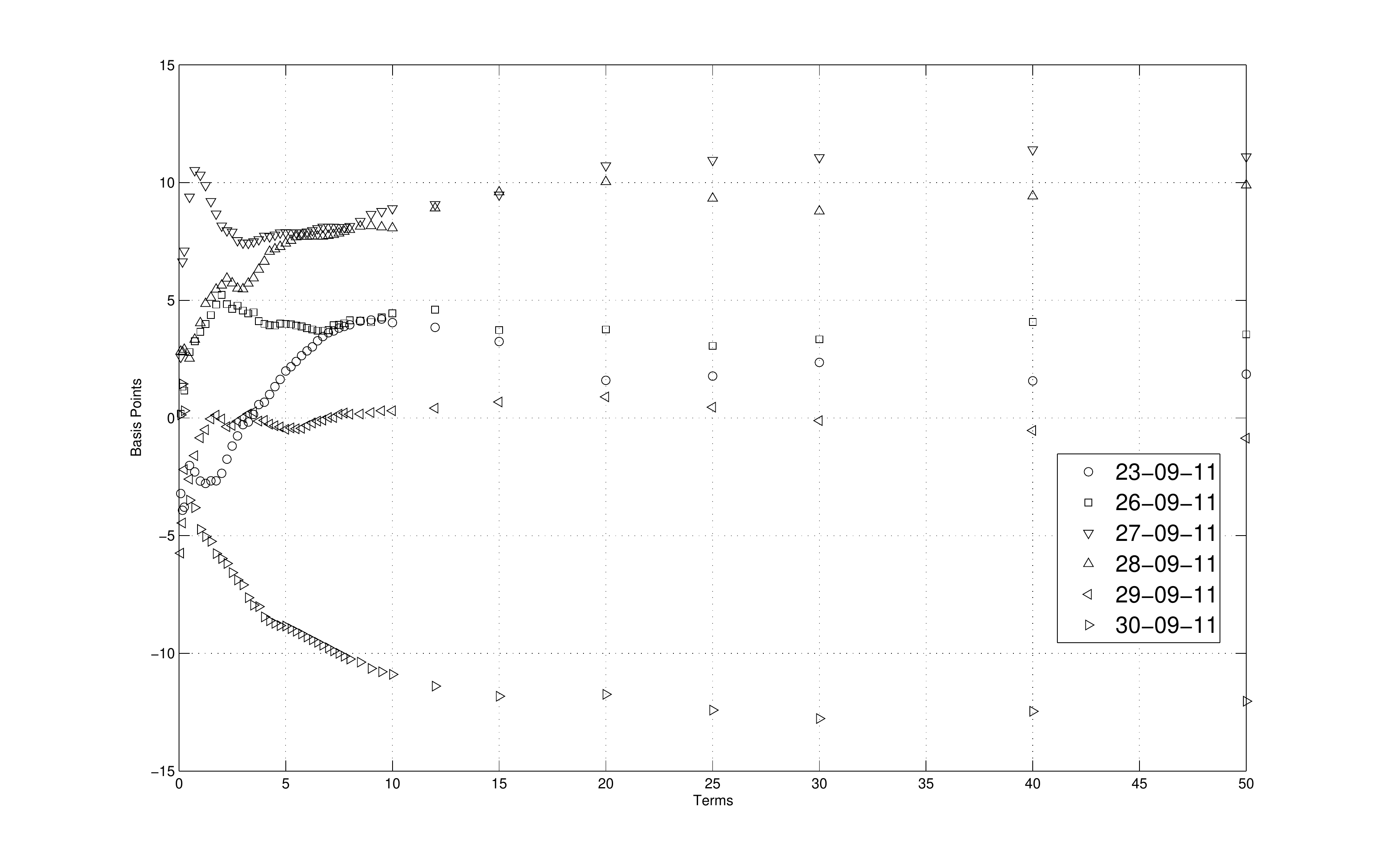}
    \label{fig:movements}
\end{figure}

The data we presented in Section (\ref{sec:DATA}) will allow us to demonstrate the effectiveness of
our algorithm to handle the stability issues.
In Figure (\ref{fig:movements}) we show the daily absolute movement of the term structure of yields. These
movements are not very small. In some cases they seem to be reducing the size of the first acute hump.
Without any stability measure, these data may generate completely different solutions.

Our approach is to modify the GA by allowing one to selectively infuse a set of genes in the initial
population. We choose these to be the winners of a previous GA run. From Table (\ref{table:parameters})
we see that the number of such genes returning from one run to another is 64 for our experiments.
The movement of yield rates is highly variable from day to day, as seen in Figure (\ref{fig:movements}).
As such, we cannot test for any pattern in the movement of the parameters. The best we can do is to check
that they do not move drastically. For the stability test, we introduce another feature to our algorithm. We
run the first experiment again for a minimum number of 10,000 generations. We then compensate for this
extra time by running the subsequent data for only 1,000 generations. The parameters and the errors
for these runs are presented in Table (\ref{table:parameterstime}).

\begin{table}[tbp]
\caption{ Daily Changes in NSS Model Parameters for the Period September 22-30, 2011}
\centering
\begin{tabular}{!{\VRule} c !{\VRule} c !{\VRule} c !{\VRule} c !{\VRule} c !{\VRule} c !{\VRule} c !{\VRule} c !{\VRule}}
\hline
$\beta_0$ & $\beta_1$ & $\beta_2$ & $\beta_3$ & $\lambda$ & $\kappa$ & $l_2$ Error & $l_\infty$ Error \\
\hline
0.020780	&	-0.011995	&	-0.034771	&	0.023232	&	1.484620	&	9.050420	&	0.001950	&	0.000942	\\
0.020662	&	-0.012027	&	-0.035353	&	0.023943	&	1.405530	&	9.186470	&	0.001921	&	0.000908	\\
0.021055	&	-0.012224	&	-0.035449	&	0.024367	&	1.430270	&	8.800990	&	0.001910	&	0.000852	\\
0.022008	&	-0.012384	&	-0.035328	&	0.024750	&	1.454020	&	9.090340	&	0.001906	&	0.000824	\\
0.022405	&	-0.012521	&	-0.034934	&	0.026274	&	1.430930	&	9.459170	&	0.002099	&	0.000872	\\
0.022416	&	-0.013029	&	-0.034438	&	0.027155	&	1.497920	&	9.110210	&	0.002057	&	0.000844	\\
0.021804	&	-0.012981	&	-0.034442	&	0.027327	&	1.654630	&	8.129460	&	0.001763	&	0.000833	\\
\hline
\end{tabular}
\label{table:parameterstime}
\end{table}

\begin{figure}[tbp]
    \centering
    \includegraphics[width=0.9\textwidth]{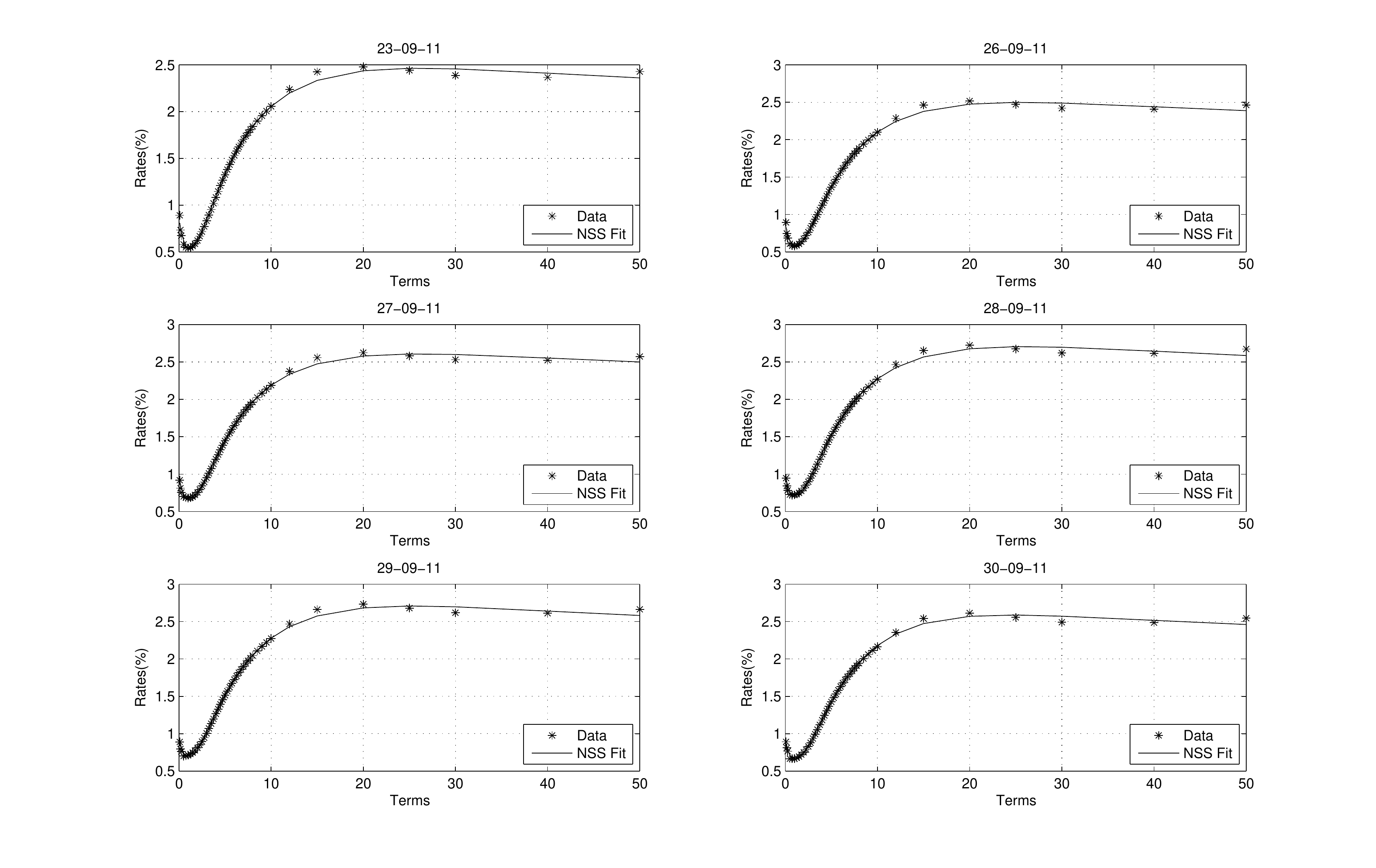}
    \caption{Fitting the OIS Curves from the Period September 22-30, 2011}
    \label{fig:nsstimes}
\end{figure}

Table (\ref{table:parameterstime}) also displays the tracking error in the $2$-norm and
the $\infty$-norm in columns 7 and 8 respectively. It is gratifying to note
that the errors in the subsequent runs are much less even with only 1,000 generations.
Not only that, the last row shows a very good fit considering the fact that
the movements are greatest for the last day in the period under study.
In practice, one could incorporate a logic where the number of generations depends on some
way on the deviation of the term structure from one day to another.
One could simply run a full 10,000 generations every week or every other week.
We do not investigate this idea here as the results show that tracking error is
second lowest on the day of maximum deviation.
We shall revisit these results in Section (\ref{sec:Conclusions}).
In Figure (\ref{fig:nsstimes}) we show how well the NSS model fits over time.
The results are visually impressive.

We conclude this Section by reporting the results of our fit on the raw data
described in Table (\ref{table:usdRates}).  $\beta_1$, $\beta_2$, and $\beta_3$
control the slope of the term structure and the degree of
curvature.  Compared with the OIS data, which has been smoothed by the cubic
spline interpolation method, the USD yield curve is harder to capture. As
such, we employ a wider range of parameter boundaries in Table
(\ref{table:usdBounds}) for this test.

\begin{table}[tbp]
\caption{Bounds Constraints for the NSS Model - USD Yield Curve}
\centering
\begin{tabular}{!{\VRule} c !{\VRule} c !{\VRule} c !{\VRule} }
\hline
Parameter & Lower Bound & Upper Bound \\
\hline
$\beta_0$ & 0.0 & 0.10 \\
$\beta_1$ & -1.0 & 4.0 \\
$\beta_2$ & -2.0 & 4.0 \\
$\beta_3$ & -2.0 & 8.0 \\
$\lambda$ & 0.0 & 6.0 \\
$\kappa$ & 6.0 & 30.0 \\
\hline
\end{tabular}
\label{table:usdBounds}
\end{table}

\begin{figure}[tbp]
    \centering
    \includegraphics[width=0.9\textwidth]{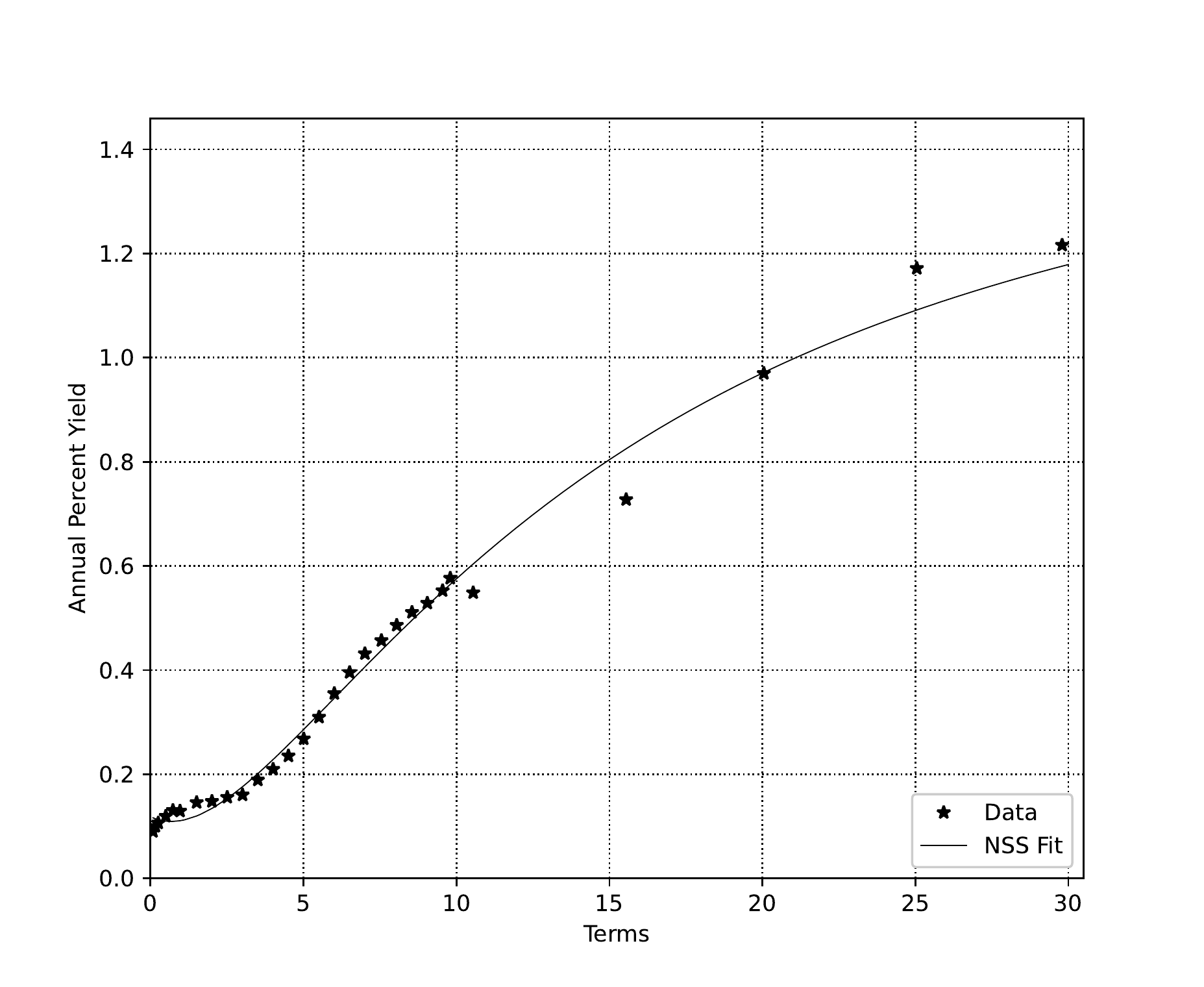}
    \caption{Fitting the USD Bond Data}
    \label{fig:usdfit}
\end{figure}

The NSS fit to raw data appears in Figure (\ref{fig:usdfit}).
We see that the raw data is fitted quite well and observe no overfitting.

\section{Conclusions and Future Research}
\label{sec:Conclusions}
We devoted the whole of Section (\ref{sec:CALIBRES}) to demonstrating the flexibility and the power of GA in dealing
with the problems associated with the calibration of the NSS model. However, some objectivity in this regard is warranted.
One is able to argue that the maximum error of about 10bp is rather high,
considering the fact that daily movements are generally smaller than 10bp.
One can see in Figures (\ref{fig:nssB}) and (\ref{fig:nsstimes}) that the third turning point is not well-captured.
The algorithm tries to fit it in the least squares sense, resulting in
prominent (point to point) tracking errors.
Nevertheless, we do not believe that any alternate algorithm will be able to
capture this region (along with the whole curve) significantly better. This is a result of the
limitation of the NSS model itself.  Secondly, fitting to a set of 45 term points is challenging.
The fitness function is a simple squared distance metric based on a set of 45 points. Loosely speaking,
the algorithm tries to capture data spread over 45 term nodes with a parametric representation with only
6 degrees of freedom. Every algorithm is susceptible to a tracking error of some magnitude.
Our approach seems to have handled the raw data effectively, without overfitting.

Having said the above, there is always a room for improvement.
Possible avenues of future research to improving the basic GA include:
\begin{itemize}
\item{Fitness Function : } One can try a different fitness function.
	For example, if one needs to capture the turning points accurately,
	one can use a weighted squared distance metric,
	allowing greater weights around the turning points.
\item{Hybrid Approach : } In practice, one cannot run the GA algorithm indefinitely,
hence convergence is an issue.  In such cases using a hybrid approach can be rewarding.
In this case a GA run can be subsequently be followed by,
for example, a non-heuristic algorithm to finely pinpoint the accurate solution.
\end{itemize}
These are just a few suggestions. This does not necessarily mean that one would obtain a considerably
better solution than the one proposed in this paper. By \textit{considerably better}, we mean one
that is able to reduce the tracking error to be insignificant when compared to the daily moves in yields.
Clearly, the model itself may not allow this possibility.

All in all, our work demonstrates the potential of GA to solve calibration issues mentioned
in Section (\ref{sec:Introduction}).  We back our claims by a series of experiments that check the
quality of fit as well as the stability of the calibrated parameter set.

\section{Acknowledgements}
\label{sec:ack}
We thank our colleagues Anastasia Aguinaldo and Hongmei Shi for providing the valuable data for our research,
and Shengjie Jin for discussions on term structure modelling.

\bibliographystyle{plainnat}

\begin{thebibliography}{36}
\providecommand{\natexlab}[1]{#1}
\providecommand{\url}[1]{\texttt{#1}}
\expandafter\ifx\csname urlstyle\endcsname\relax
  \providecommand{\doi}[1]{doi: #1}\else
  \providecommand{\doi}{doi: \begingroup \urlstyle{rm}\Url}\fi

\bibitem[Adams and Deventer(1994)]{adams1994}
K.~J. Adams and D.~R.~Van Deventer.
\newblock {F}itting yield curves and forward rate curves with maximum
  smoothness.
\newblock \emph{Journal of Fixed Income}, 4\penalty0 (1):\penalty0 52--62,
  1994.

\bibitem[Anderson and Sleath(2001)]{boe2001}
N.~Anderson and J.~Sleath.
\newblock {N}ew estimates of the {UK} real and nominal yield curves.
\newblock \emph{Bank of {E}ngland Quarterly Bulletin}, 2001.

\bibitem[Annaert et~al.(2013)Annaert, Claes, {De Ceuster}, and
  Zhang]{annaert2013}
J.~Annaert, A.~G.P. Claes, M.~J.K. {De Ceuster}, and H.~Zhang.
\newblock {E}stimating the yield curve using the {N}elson-{S}iegel model, a
  ridge regression approach.
\newblock \emph{International Review of Economics and Finance}, 2013.

\bibitem[Barrett et~al.(1995)Barrett, Gosnell, and Heuson]{barrett1995}
W.~R. Barrett, T.~F.~Jr. Gosnell, and A.~J. Heuson.
\newblock {Y}ield curve shifts and the selection of immunization strategies.
\newblock \emph{Journal of Fixed Income}, 5\penalty0 (2):\penalty0 53--64,
  1995.

\bibitem[{BRODA}(2016)]{broda2016}
{BRODA}.
\newblock Broda - software.
\newblock \url{http://www.broda.co.uk/software.html}, 2016.
\newblock [Online; accessed Dec 12, 2017].

\bibitem[Cabrera et~al.(2014)Cabrera, Lammers, d.~P.~Morón, and
  Vega]{cabrera2014}
D.~Cabrera, J.~H. Lammers, M.~d.~P.~Morón, and A.~Vega.
\newblock {T}he {N}elson-{S}eigel-{S}vensson approach.
\newblock 2014.

\bibitem[Cairns and Pritchard(2001)]{cairns2001}
A.~J.~G. Cairns and D.~J. Pritchard.
\newblock {S}tability of descriptive models for the term structure of interest
  rates with applications to {G}erman market data.
\newblock \emph{{B}ritish Actuarial Journal}, 7:\penalty0 467--507, 2001.

\bibitem[Coroneo et~al.(2011)Coroneo, Nyholm, and Vivada-Koleva]{coroneo2011}
L.~Coroneo, K.~Nyholm, and R.~Vivada-Koleva.
\newblock {H}ow arbitrage-free is the {N}elson-{S}iegel model?
\newblock \emph{Journal of Empirical Finance}, 18:\penalty0 393--407, 2011.

\bibitem[{De Pooter}(2007)]{dePooter2007}
M.~{De Pooter}.
\newblock {E}xaming the {N}elson-{S}iegel class of term structure models.
\newblock \emph{{T}inbergen Institute Discussion Paper}, 2007.

\bibitem[Diebold and Li(2006)]{diebold2006}
F.~X. Diebold and C.~Li.
\newblock {F}orecasting the term structure of government bond yields.
\newblock \emph{Journal of Econometrics}, 130:\penalty0 337--364, 2006.

\bibitem[Diebold and Rudebusch(2013)]{diebold2013}
F.~X. Diebold and G.~D. Rudebusch.
\newblock \emph{{Y}ield Curve Modeling and Forecasting: The Dynamic
  {N}elson-{S}iegel Approach}.
\newblock {P}rinceton University Press, 2013.
\newblock ISBN 9781400845415.

\bibitem[Fabozzi et~al.(2005)Fabozzi, Martellini, and Priaulet]{fabozzi2005}
F.~J. Fabozzi, L.~Martellini, and P.~Priaulet.
\newblock {P}redictability in the shape of the term structure of interest
  rates.
\newblock \emph{Journal of Fixed Income}, 15\penalty0 (1):\penalty0 40--53,
  2005.

\bibitem[Fisher et~al.(1994)Fisher, Nychka, and Zervous]{fisher1994}
M.~Fisher, D.~Nychka, and D.~Zervous.
\newblock {F}itting the term structure of interest rates with smoothing
  splines.
\newblock Technical report, 1994.

\bibitem[Gilli and Schumann(2010)]{gilliSchumann2010}
M.~Gilli and E.~Schumann.
\newblock {A} note on ‘good starting values’ in numerical optimisation.
\newblock \emph{{COMISEF} working paper series}, \penalty0 (44), 2010.

\bibitem[Gilli et~al.(2010)Gilli, Große, and
  Schumann]{gilliGrosseSchumann2010}
M.~Gilli, S.~Große, and E.~Schumann.
\newblock {C}alibrating the {N}elson-{S}iegel-{S}vensson model.
\newblock \emph{{COMISEF} working paper series}, \penalty0 (31), 2010.

\bibitem[Goldberg(1989)]{goldberg1989}
D.~E. Goldberg.
\newblock \emph{{G}enetic Algorithms in Search, Optimization \& Machine
  Learning}.
\newblock {A}ddison-{W}esley, 1 edition, 1989.
\newblock ISBN 0-201-15767-5.

\bibitem[Gurkaynak et~al.(2007)Gurkaynak, Sack, and Wright]{gurkaynak2007}
R.~S. Gurkaynak, B.~Sack, and J.~H. Wright.
\newblock {T}he {U}.{S}. treasury yield curve: 1961 to the present.
\newblock \emph{Journal of Monetary Economics}, 54\penalty0 (8):\penalty0
  2291--2304, 2007.

\bibitem[Haupt and Haupt(2004)]{haupt2004}
Randy~L. Haupt and Sue~Ellen Haupt.
\newblock \emph{{P}ractical Genetic Algorithms}.
\newblock {W}iley, 2 edition, 2004.
\newblock ISBN 0-471-45565-3.

\bibitem[Hull and White(2012)]{hull2012}
John Hull and Allan White.
\newblock {LIBOR} vs. {OIS}: The derivatives discounting dilemma.
\newblock \emph{{J}ournal Of Investment Management}, 11\penalty0 (3):\penalty0
  14--27, 2012.

\bibitem[Lakhany and Mausser(2000)]{lakhany2000}
Asif Lakhany and Helmut Mausser.
\newblock {E}stimating the parameters of the generalized lambda distribution.
\newblock \emph{{A}lgo Research Quarterly Journal}, 3\penalty0 (3), 2000.

\bibitem[Litterman and Scheinkman(1991)]{litterman1991}
R.~Litterman and J.~Scheinkman.
\newblock {C}ommon factors affecting bond returns.
\newblock \emph{Journal of Fixed Income}, 1\penalty0 (1):\penalty0 54--61,
  1991.

\bibitem[Lorenčič(2016)]{lorencic2016}
E.~Lorenčič.
\newblock {T}esting the performance of cubic splines and {N}elson-{S}iegel
  model for estimating the zero-coupon yield curve.
\newblock \emph{Journal of Contemporary Issues in Economics and Business},
  62\penalty0 (2):\penalty0 42--50, 2016.

\bibitem[Martellini and Meyfredi(2007)]{martellini2007}
L.~Martellini and J.~C. Meyfredi.
\newblock {A} copula approach to value-at-risk estimation for fixed-income
  portfolio.
\newblock \emph{Journal of Fixed Income}, 17\penalty0 (1):\penalty0 5--15,
  2007.

\bibitem[McCulloch(1971)]{mcculloch1971}
J.~H. McCulloch.
\newblock {M}easuring the term structure of interest rates.
\newblock \emph{Journal of Business}, 44:\penalty0 19--31, 1971.

\bibitem[McCulloch(1975)]{mcculloch1975}
J.~H. McCulloch.
\newblock {T}he tax-adjusted yield curve.
\newblock \emph{Journal of Finance}, 30:\penalty0 811--830, 1975.

\bibitem[{Monetary \& Economic Department}(2005)]{bis25}
{Monetary \& Economic Department}.
\newblock {Z}ero-coupon yield curves estimated by central banks.
\newblock Technical Report~25, Bank for International Settlements, 2005.
\newblock URL \url{https://www.bis.org/publ/bppdf/bispap25a.pdf}.

\bibitem[Nelson and Siegel(1987)]{nelson1987}
C.~R. Nelson and A.~F. Siegel.
\newblock {P}arsimonious modelling of yield curves.
\newblock \emph{Journal of Business}, 60:\penalty0 473--489, 1987.

\bibitem[Nielsen(2015)]{nielsen2017}
Barry Nielsen.
\newblock {B}ond yield curve holds predictive powers.
\newblock 2015.
\newblock URL
  \url{https://www.investopedia.com/articles/economics/08/yield-curve.asp}.

\bibitem[Seber and Wild(2003)]{seber2003}
G.~A.~F. Seber and C.~J. Wild.
\newblock {N}onlinear regression.
\newblock \emph{{W}iley Series in Probability and Statistics}, 2003.

\bibitem[Shea(1984)]{shea1984}
G.~S. Shea.
\newblock {P}itfalls in smoothing interest rate term structure data:
  Equilibrium models and spline approximations.
\newblock \emph{Journal of Financial and Quantitative Analysis}, 19:\penalty0
  253--269, 1984.

\bibitem[Steeley(1991)]{steeley1991}
J.~M. Steeley.
\newblock {E}stimating the {G}ilt-edged term structure: Basis splines and
  confidence intervals.
\newblock \emph{Journal of Business Finance \& Accounting}, 18\penalty0
  (4):\penalty0 513--529, 1991.

\bibitem[Svensson(1994)]{svensson1994}
L.~Svensson.
\newblock {E}stimating and interpreting forward interest rates: {S}weden
  1992-94.
\newblock \emph{{IMF} working paper series}, \penalty0 (114), 1994.

\bibitem[Svensson(1995)]{svensson1995}
L.~Svensson.
\newblock {E}stimating forward interest rates with the extended {N}elson \&
  {S}iegel model.
\newblock \emph{{S}veriges Riksbank Quarterly Review}, \penalty0 (3):\penalty0
  13--26, 1995.

\bibitem[Vasicek and Fong(1982)]{vasicek1982}
O.~A. Vasicek and H.~G. Fong.
\newblock {T}erm structure modelling using exponential splines.
\newblock \emph{Journal of Finance}, 73:\penalty0 339--348, 1982.

\bibitem[Waggoner(1997)]{waggoner1997}
D.~Waggoner.
\newblock {S}pline methods for extracting interest rate curves from coupon bond
  prices.
\newblock Technical report, Federal Reserve Bank of {A}tlanta, 1997.

\bibitem[Xie et~al.(2008)Xie, Wu, and Shi]{xie2008}
Y.~A. Xie, C.~Wu, and J.~Shi.
\newblock {D}o macroeconomic variables matter for the pricing of default risk?
  {E}vidence from the residual analysis of the reduced-form model pricing
  errors.
\newblock \emph{International Review of Economics \& Finance}, 17\penalty0
  (2):\penalty0 279--291, 2008.

\end{thebibliography}

\end{document}